\documentstyle [12pt]{article}
\begin{document}
\newcommand {\beq}{\begin{equation}}
\newcommand {\eeq}{\end{equation}}
\newcommand {\beqa}{\begin{eqnarray}}
\newcommand {\eeqa}{\end{eqnarray}}
\newcommand {\sfrac}[2]{{\textstyle \frac{#1}{#2}}}
\newcommand {\n}{\nonumber \\}
\newcommand {\eqn}[1]{(\ref{#1})}
\newcommand {\eq}[1]{eq.~(\ref{#1})}
\newcommand {\eqs}[1]{eqs.~(\ref{#1})}
\newcommand {\Eq}[1]{Eq.~(\ref{#1})}
\newcommand {\eql}[1]{eqs.~(\ref{#1},}
\newcommand {\eqr}[1]{\ref{#1})}
\newcommand {\eqm}[1]{\ref{#1},}
\newcommand {\Label}[1]{\label{#1}}
\newcommand {\Bibitem}[1]{\bibitem{#1}}
\begin{flushright}
{\boldmath $Y\!ukawa$ $Institute$ $Kyoto$}\hfill
YITP/K-987\\
August 1992\\
hep-th/9209019
\end{flushright}
\vskip 2em
\begin{center}
{\LARGE Continuum Limit of Spin-1 Chain \par} \vspace{25mm}
{\large \lineskip .5em
Takeo Inami 
and
Satoru Odake\footnote{Fellow of Soryushi Shogakukai.}
\par} \vspace{15mm}
{\large {\sl Yukawa Institute for Theoretical Physics\\
              Kyoto University, Kyoto 606, Japan}}
\end{center} \par
\vspace{20mm}
\begin{abstract}
We study the continuum limit of the spin-1 chain in the non-Abelian
bosonization approach of Affleck and show that the Hamiltonian of
integrable spin-1 chain yields the Lagrangian of supersymmetric
sine-Gordon model in the zero lattice spacing limit. We also show
that the quantum group generators of the spin-1 chain give non-local
charges of the supersymmetric sine-Gordon theory.
\end{abstract}

\vspace{10mm}

\newpage
One-dimensional quantum spin chains have played a very important
role in the theory of highly correlated electron systems. There
has recently been experimental interest in quasi one-dimensional
systems of loosely coupled molecules. They can be described
(approximately) by anti-ferromagnetic spin chains. Some of them
have spin $s$ greater than $\frac{1}{2}$.
Spin chains have also played a special role in the theory
of integrable systems. XXZ spin chains are known to have
quantum group symmetry $U_qsu(2)$ \cite{PaSa}.
This symmetry underlies the integrability of spin chains.
Quantum group symmetries\cite{JiDr}
also appear in some class of quantum field theories in $1+1$
dimensions, e.g. non-local conserved charges in the sine-Gordon
theory generate  the affine $U_q\widehat{su}(2)$ symmetry\cite{BeLe}.
In fact, the
continuum (field theory) limit of the XXZ spin-$\frac{1}{2}$
chain is known to be the sine-Gordon theory\cite{LuLudN,Af'}.

In this letter we consider the continuum limit of the spin-1 chain
in the non-Abelian bosonization approach of Affleck\cite{Af}.
We show that the Hamiltonian of the integrable spin-1 chain
yields the Lagrangian of supersymmetric sine-Gordon model in
the zero lattice spacing limit and that
the $U_q\widehat{su}(2)$ quantum group generators of the
infinitely long chain give non-local charges of the latter theory.

It is believed that, in contrast to the spin-$\frac{1}{2}$ case, the
spin-1 Heisenberg model has a mass gap\cite{Ha}. The
more general spin-1 chain with isotropic bilinear-biquadratic
Hamiltonian
$ H=J\sum_n [\vec{S}_{n+1}\cdot\vec{S}_n
  -b(\vec{S}_{n+1}\cdot\vec{S}_n)^2]$
is also conjectured to be gapful for $b\neq \pm 1$.
We are interested in the case of $b=1$, which is an integrable point
and gapless\cite{KuSkTaBa},
and hence it makes sense to consider its field theory
limit. Integrable deformation of this Hamiltonian
incorporating anisotropy was constructed in ref.\cite{ZaFa}
and was further studied in ref.\cite{DSZ}.
\beqa
 H_{\!X\!X\!Z}&\!\!=\!\!&J\sum_n \Bigl[
    \vec{S}_{n+1}\cdot\vec{S}_n
    -(\vec{S}_{n+1}\cdot\vec{S}_n)^2 \n
 && +\sfrac{1}{2}(q-q^{-1})^2
    (S^3_{n+1}S^3_n-(S^3_{n+1}S^3_n)^2+(S^3_{n+1})^2+(S^3_n)^2) \n
 && -\sfrac{1}{2}(q+q^{-1}-2)
    \lbrace S^3_{n+1}S^3_n,S^+_{n+1}S^-_n+S^-_{n+1}S^+_n \rbrace
    +\sfrac{1}{2}(q^2-q^{-2})(S^3_{n+1}-S^3_{n})
    \Bigr].
 \Label{XXZ}
\eeqa
This XXZ Hamiltonian commutes with $U_qsu(2)$, whose generators
are $H_1$ and $E^{\pm}_1$:
\beq
 q^{H_1}=\cdots q^{2S^3_{n+1}}q^{2S^3_n}q^{2S^3_{n-1}} \cdots,~~~
 E^{\pm}_1=\sqrt{[2]/2}\sum_n\cdots q^{S^3_{n+1}}
    S^{\pm}_n q^{-S^3_{n-1}}\cdots,
 \Label{Uq}
\eeq
where $S^{\pm}=S^1\pm iS^2$ and $[x]=(q^x-q^{-x})/(q-q^{-1})$.
For infinitely long chain, in which case the last
boundary term can be discarded,
this symmetry is enhanced to the affine quantum group symmetry
$U_q\widehat{su}(2)$ with level 0 \cite{Da}, whose generators are $H_1$,
$E^{\pm}_1$, $H_0=-H_1$ and
$ E^{\pm}_0=\sqrt{[2]/2}\sum_n\cdots q^{-S^3_{n+1}}
    S^{\mp}_n q^{S^3_{n-1}}\cdots$.

To derive the continuum limit of \eqn{XXZ} we use the oscillator
representation of spin-1 operator $S^a_n$. In view of the lack
of the Jordan-Wigner transformation for the spin-1 case, we employ
the construction of $S^a_n$ out of a pair of
spin-$\frac{1}{2}$ operators\cite{Af,Af'}.
We will comment on other methods later.
To get $s=1$ we need two ($f=1,2$) replicas of doublet ($\alpha=1,2$)
complex fermions $\psi^{\alpha f}_n$. They obey the
canonical anticommutation relations
$\lbrace \psi^{\alpha f}_n,\psi^{\beta g\dagger}_m \rbrace =
\delta^{\alpha\beta}\delta^{fg}\delta_{nm}$. The spin-1 operator
is represented as
\beq
 S^a_n=\sfrac{1}{2}\psi^{\alpha f\dagger}_n\sigma^a_{\alpha\beta}
       \psi^{\beta f}_n,
 \Label{Sa}
\eeq
where $\sigma^a$'s are Pauli matrices.
On each lattice site there are $2^4$ states obtained by acting creation
operators $\psi^{\alpha f\dagger}_n$ on the vacuum $|0\rangle_n$ defined
by $\psi^{\alpha f}_n|0\rangle_n=0$. Spin-1 states $|1,m\rangle_n$ are
obtained by acting the lowering operator $S^-_n$ on the highest weight
state $|1,1\rangle_n=\prod_{f}\psi^{1f\dagger}_n|0\rangle_n$.
This spin-1 representation space is characterized by\cite{Af}
\beq
 \vec{S}^2_n|*\rangle^L=2|*\rangle^L,~~~\mbox{or equivalently}~~~
 \psi^{\alpha f\dagger}_n\psi^{\alpha g}_n|*\rangle^L
 =\delta^{fg}|*\rangle^L.
 \Label{const}
\eeq
Here the suffix $L$ refers to the lattice theory.
Thus we impose the above constraint to project out three spin-1 states
$|1,m\rangle_n$ from 16 states on each site.

The second equation of \eqn{const} means that a half of the
particle states is filled on each site. In the case of
spin-$\frac{1}{2}$, an analysis of the Hubbard model shows that
this condition implies half-filling in the momentum space.
By a similar analysis of multi-band Hubbard models,
one can argue that this is the case for higher spins\cite{Af,Af'}.
We define the half-filling
vacuum $|0\rangle^{HF}$ as the state in which particle
states are filled up to the Fermi sea of momentum $k_F=\pi/2a$.
Low energy excitations are creations of fermions and holes near the
Fermi sea. To describe such excitations we introduce chiral fermions
$\psi^{\alpha f}_{\pm,\ell}$ on a pair of even and odd lattice sites:
\beq
 \psi^{\alpha f}_{2\ell}=(-1)^\ell
     (\psi^{\alpha f}_{+,\ell}+\psi^{\alpha f}_{-,\ell})/\sqrt{2},~~
 \psi^{\alpha f}_{2\ell-1}=-i(-1)^\ell
     (\psi^{\alpha f}_{+,\ell}-\psi^{\alpha f}_{-,\ell})/\sqrt{2}.
\eeq
The fast oscillating term $(-1)^{\ell}$ comes from
$\exp(\pm ik_Fa2\ell)$ and $\exp(\pm ik_Fa(2\ell-1))$.
The spin operator \eqn{Sa} is expressed in terms of chiral fermions as
\beq
 S^a_{2\ell}=\sfrac{1}{2}(J^a_{\ell}+G^a_{\ell}),~~~
 S^a_{2\ell-1}=\sfrac{1}{2}(J^a_{\ell}-G^a_{\ell}),
\eeq
where
\beq
 J^a_{\pm,\ell}=
   \sfrac{1}{2}\psi^{\alpha f\dagger}_{\pm,\ell}\sigma^a_{\alpha\beta}
   \psi^{\beta f}_{\pm,\ell},~~~
 J^a_{\ell}=J^a_{+,\ell}+J^a_{-,\ell},~~~
 G^a_{\ell}=
   \sfrac{1}{2}(\psi^{\alpha f\dagger}_{+,\ell}\sigma^a_{\alpha\beta}
                \psi^{\beta f}_{-,\ell}
              +\psi^{\alpha f\dagger}_{-,\ell}\sigma^a_{\alpha\beta}
               \psi^{\beta f}_{+,\ell}).
\eeq

We will present the derivation of the continuum limit of the
Hamiltonian $H$ of isotropic spin-1 chain. Extension to the
anisotropic case $H_{\!X\!X\!Z}$ will be briefly discussed later.
We take the zero lattice spacing limit ($a\rightarrow 0$).
The space coordinate is $x=2a\ell$ and the sum is replaced by
the integral $2a\sum_{\ell}\rightarrow \int dx$.
There are a few alternative ways of computing the continuum
Hamiltonian depending on the different stages at which we move from
the lattice to continuum theory. We take the prescription of
taking the $a\rightarrow 0$ limit in an early stage and computing
the operator products of currents in the continuum theory.
We have also made the computation in the lattice theory taking
the $a\rightarrow 0$ limit in the resulting expression. We have
obtained the same physical results (modulo some subtleties related to
regularization).

In the continuum
$ \frac{1}{\sqrt{2a}}\psi^{\alpha f}_{\pm,\ell}\rightarrow
  \psi^{\alpha f}_{\pm}(x)$
and their propagators are
$ \langle \psi^{\alpha f\dagger}_{\pm}(x)\psi^{\beta g}_{\pm}(y)\rangle
  =
  \delta^{\alpha\beta}\delta^{fg}
  (\mp 2\pi i)^{-1}(x-y\pm i\epsilon)^{-1}$,
where $\pm i\epsilon$ is the UV cutoff in the continuum theory.
We assume the existence of the continuum (field theory) limit of
lattice spin-1 states and the half-filled vacuum:
$|*\rangle^L\rightarrow |*\rangle^{FT}$,
$|0\rangle^{HF}\rightarrow |0\rangle^{FT}$.
Here suffices $L$, $FT$ and $HF$ refer to lattice, field theory and
half filling.
Normal ordering of fermions refers to this $|0\rangle^{FT}$.
Currents in the continuum are
\beqa
 (2a)^{-1}J^a_{\pm,\ell}&\rightarrow& J^a_{\pm}(x)=
   \sfrac{1}{2}(\psi^{\alpha f\dagger}_{\pm}\sigma^a_{\alpha\beta}
              \psi^{\beta f}_{\pm})(x),\\
 (2a)^{-1}G^a_{\ell}&\rightarrow& G^a(x)=
   \sfrac{1}{2}(\psi^{\alpha f\dagger}_+\sigma^a_{\alpha\beta}
               \psi^{\beta f}_-
               +\psi^{\alpha f\dagger}_-\sigma^a_{\alpha\beta}
               \psi^{\beta f}_+)(x)
\eeqa
and their operator product expansions (OPE) are easily calculated.
For example,
\beqa
 J^a_{\pm}(x) J^b_{\pm}(0)&\!\!=\!\!&(\mp 2\pi ix)^{-2}\delta^{ab}
     +(\mp 2\pi ix)^{-1}i\epsilon^{abc}J^c_{\pm}(0)
     +(J^a_{\pm}J^b_{\pm})(0)+\cdots,
 \Label{su2} \\
 J^a_{\pm}(x) G^b(0)&\!\!=\!\!&
     (\mp 2\pi ix)^{-1}(\pm\sfrac{1}{4}\delta^{ab}F(0)
                        +\sfrac{1}{2}i\epsilon^{abc}G^c_{\pm}(0))
     +(J^a_{\pm}G^b)(0)+\cdots,
 \Label{JG}
\eeqa
where
$ F(x)=(\psi^{\alpha f\dagger}_+\psi^{\alpha f}_-
       -\psi^{\alpha f\dagger}_-\psi^{\alpha f}_+)(x)$.
The first equation
means that $J^a_{\pm}$ define the $\widehat{su}(2)\times
\widehat{su}(2)$ Kac-Moody algebra of level $k=2$, as we have
designed. Following Affleck we assume that the states
in the field theory satisfy the continuum limit
of the spin-1 constraint \eqn{const}\cite{Af}
\beq
 (\vec{J}^{~2}+\vec{G}^2)(x)|*\rangle^{FT}=0,~~~
 (\vec{J}\cdot\vec{G}+\vec{G}\cdot\vec{J})(x)|*\rangle^{FT}=0,
 \Label{const2}
\eeq
where $(AB)(x)$ stands for normal ordering defined by the regular
part of OPE.

We are now ready to compute the $a\rightarrow 0$ limit of the
Hamiltonian $H$.
Using $(\vec{S}_{n+1}\cdot\vec{S}_n)^2=
\frac{1}{4}\lbrace S^a_{n+1},S^b_{n+1}\rbrace
\lbrace S^a_n,S^b_n\rbrace-\frac{1}{2}\vec{S}_{n+1}\cdot\vec{S}_n$,
the Hamiltonian is now written as
\beq
 H=2aJ\int dx \Bigl[
   \sfrac{3}{2}({\cal H}^{(2)}_e+{\cal H}^{(2)}_o)(x)
   -({\cal H}^{(4)}_e+{\cal H}^{(4)}_o)(x)\Bigr],
\eeq
where suffices $e$ and $o$ refer to even and odd $n$, and
\beqa
 &&(2a)^{-2} \vec{S}_{2\ell+1}\cdot\vec{S}_{2\ell} ~~\rightarrow ~~
 {\cal H}^{(2)}_e(x)=\sfrac{1}{4}
  (\vec{J}-\vec{G})(x+2a)\cdot (\vec{J}+\vec{G})(x), \\
 &&(2a)^{-2} \sfrac{1}{4}\lbrace S^a_{2\ell+1},S^b_{2\ell+1}\rbrace
 \lbrace S^a_{2\ell},S^b_{2\ell}\rbrace ~~
 \rightarrow \n
 &&{\cal H}^{(4)}_e(x)=(2a)^2\sfrac{1}{16}
 [(\pi\epsilon)^{-2}\delta^{ab}+(J^aJ^b+G^aG^b)(x+2a)
                               -(J^aG^b+G^aJ^b)(x+2a)] \n
 &&\hspace{29mm}
 \times[(\pi\epsilon)^{-2}\delta^{ab}+(J^aJ^b+G^aG^b)(x)
                                     +(J^aG^b+G^aJ^b)(x)],
\eeqa
and similar expressions for ${\cal H}^{(2)}_o$ and ${\cal H}^{(4)}_o$.
We have used the fact that $(J^aJ^b+G^aG^b)$ and
$(J^aG^b+G^aJ^b)$ are symmetric in $a$ and $b$.
The relevant terms of the Hamiltonian can be obtained using
the OPE such as \eqn{su2} and \eqn{JG}. The results are
\beqa
 {\cal H}^{(2)}_e+{\cal H}^{(2)}_o&\!\!=\!\!&
 \sfrac{1}{2}(\vec{J}^{~2}-\vec{G}^2), \\
 {\cal H}^{(4)}_e+{\cal H}^{(4)}_o&\!\!=\!\!&
 -1/(2\pi)^2[\sfrac{1}{2}(\vec{J}^{~2}-\vec{G}^2)
   +\sfrac{1}{2}(\vec{K}^2)+\sfrac{3}{4}(F^2)] \n
 &&
 +(a/\pi\epsilon)^2\sfrac{1}{2}(\vec{J}^{~2}+\vec{G}^2)
 -(2a)^{-1}15i/(32\pi^3)F,
\eeqa
where $\vec{K}=\vec{J}_+-\vec{J}_-$ (the coefficients depend on
how the continuum theory is regularized).
There appear operators $F$, $\vec{G}^2$
and $F^2$ in addition to the composites of the currents $J^a_{\pm}$
of the $\widehat{su}(2)$ Kac-Moody algebra of level 2, which we
denote by $\widehat{su}(2)_2$.
The divergent term $a^{-1}F$ violates the invariance under the
translation by $a$. We should discard this term assuming the lattice
regularization respecting this invariance.

Introducing time $t$, the Hamiltonian can be converted to the
Lagrangian
\beq
 {\cal L}=\sfrac{1}{2}i(
   \psi^{\alpha f\dagger}_+
   \stackrel{\leftrightarrow}{\partial}_0 \psi^{\alpha f}_+
  +\psi^{\alpha f\dagger}_-
   \stackrel{\leftrightarrow}{\partial}_0 \psi^{\alpha f}_-)
  -{\cal H}.
\eeq
As we have designed,
$\psi^{\alpha f}_{\pm}(x)$ become right(left)-moving
fermions $\psi^{\alpha f}_{\pm}(x^{\pm})$,
where $x^{\pm}=x^0\mp x^1=t\mp x$.
By evaluating the operator products of $G^a(x^+,x^-)$ and $F(x^+,x^-)$
with $J^a_{\pm}(x^{\pm})$, we have found that $G^a$ and $F$ are spin
$(\frac{1}{2},\frac{1}{2})$ multiplet of
$\widehat{su}(2)_2\times\widehat{su}(2)_2$
and $(\vec{G}^2-\frac{1}{4}F^2)$ is a singlet. This implies that we
should set
$ (F^2-4\vec{G}^2)(x)=0$.
After using this relation and the constraint \eqn{const2},
the Hamiltonian is
\beqa
 H&\!\!=\!\!&2aJ\int dx \bigl [ A(\vec{J}_+^{~2}+\vec{J}_-^{~2})(x)
     +2B(\vec{J}_+\cdot\vec{J}_-)(x) \bigr ],
 \Label{H} \\
 A&\!\!=\!\!&3[1-1/(2\pi)^2]/2,~~~
 B=[3-5/(2\pi)^2]/2.
\eeqa

We now begin to see the emergence of the supersymmetric
sine-Gordon theory. After normalizing $J$ correctly($J\sim a^{-1}$),
the first term on the r.h.s. of \eqn{H} is the Hamiltonian of the
Wess-Zumino-Witten model with level 2. This model has the central
charge $c=3/2$ and is supersymmetric. The second term is a
perturbation to the conformal invariant theory and the resulting
theory is the super sine-Gordon theory with $\beta=\sqrt{4\pi}$
\cite{ABL,KU}.

To express the Hamiltonian in a more familiar form,
we use the fact that the Kac-Moody algebra $\widehat{su}(2)_2$
is represented by a real boson and a real fermion
(${\bf Z}_2$ parafermion)\cite{ZaFa}:
\beq
 J^3_{\pm}(x^{\pm})=\pm\sqrt{1/\pi}\partial_{\pm} \phi_{\pm}(x^{\pm}),~~~
 J^+_{\pm}(x^{\pm})=\psi_{\pm}(x^{\pm})\sqrt{\mu/\pi}
        e^{\pm i\sqrt{4\pi}\phi_{\pm}(x^{\pm})},~~~
 J^-_{\pm}=(J^+_{\pm})^{\dagger},
\eeq
where we have suppressed the normal ordered symbol and
\lq\lq cocycle factors"
which ensure the commutativity of right and left currents.
The propagators of $\phi_{\pm}$ and $\psi_{\pm}$ are
$ \langle \phi_{\pm}(x^{\pm})\phi_{\pm}(0) \rangle =
  -\frac{1}{4\pi}\log(i\mu x^{\pm})$,
$ \langle \psi_{\pm}(x^{\pm})\psi_{\pm}(0) \rangle =
  (2\pi ix^{\pm})^{-1}$.
Using the identity\cite{Wi}
$\mu^2\cos^2(\sqrt{4\pi}\phi)=\frac{\pi}{2}(\partial \phi)^2$,
the Lagrangian becomes
\beq
 {\cal L}=\frac{1}{2}\partial_{\mu}\phi\partial^{\mu}\phi
 +\frac{i}{2}\bar{\psi}\gamma^{\mu}\partial_{\mu}\psi
 +\frac{1}{4}\lambda\beta\bar{\psi}\psi\cos(\beta\phi)
 +\frac{1}{8}\lambda^2\beta^2\cos^2(\beta\phi),
 \Label{ssG}
\eeq
where $\beta=\sqrt{4\pi}$ and $\lambda=\mu B/2\pi A$.
Here $\psi=(\psi_+,\psi_-)^T$.
This is the super sine-Gordon model with $\beta=\sqrt{4\pi}$.
For this value of $\beta$ the last two terms in \eqn{ssG} are
(irrelevantly) marginal.

To get the model with $\beta<\sqrt{4\pi}$ we should consider
the anisotropic case $H_{\!X\!X\!Z}$.
A new feature is that the Hamiltonian \eqn{XXZ} contains
the $(G^3G^3)$ term. In the case of spin-$\frac{1}{2}$,
the $(G^3G^3)$ term can be expressed in terms of
$\widehat{su}(2)_1$ currents $J^a_{\pm}$ after solving the constraint
like \eqn{const}. In the case of spin-1, we expect that
the $(G^3G^3)$ term
can be expressed in terms of $\widehat{su}(2)_2$ currents $J^a_{\pm}$
by solving the constraint \eqn{const}. Then, the anisotropic
terms $(J^3J^3)$ result in a change of the normalization of $\phi$ and
we must rescale $\phi\rightarrow\frac{\beta}{\sqrt{4\pi}}\phi$,
as discussed in ref.\cite{Af'} in the case of spin-$\frac{1}{2}$.
Finally we get the Lagrangian of supersymmetric sine-Gordon theory
\eqn{ssG} with a general value of $\beta$.
Unfortunately we are not yet able to solve the constraint \eqn{const}
explicitly.

Next we consider the continuum limit of the $U_q\widehat{su}(2)_0$
generators \eqn{Uq}. They are rewritten, without any approximation,
as $ q^{H_1}=q^{2\sum_{\ell}J^3_{\ell}}$ and
\beq
 E^{\pm}_1=\sqrt{[2]/2}\sum_{\ell}
  q^{\sum_{\ell'>\ell}J^3_{\ell'}}
  \Bigl[S^{\pm}_{2\ell}q^{-S^3_{2\ell-1}}
  +q^{S^3_{2\ell}}S^{\pm}_{2\ell-1}\Bigr]
  q^{-\sum_{\ell'<\ell}J^3_{\ell'}}.
\eeq
Taking the continuum limit, they are expressed in terms of
$J^a_{\pm}$ as
$ q^{H_1}=q^{2\int dxJ^3(x)}$ and
\beq
 E^{\pm}_1=\sqrt{[2]/2}\int_{-\infty}^{\infty} dx
  q^{\int^{\infty}_x dx'J^3(x')}
  J^{\pm}(x)
  q^{-\int^x_{-\infty} dx'J^3(x')},
\eeq
and similar expressions for $H_0$ and $E^{\pm}_0$.
The chiral bosons $\phi_{\pm}$ are represented in terms of $\phi$ and
its conjugate momentum $\pi=\partial_0\phi$ as
$ \phi_{\pm}(t,x)=
  \frac{1}{2}[\phi(t,x)\mp\int^x_{-\infty}dx'\pi(t,x')]$.
Under the rescaling $\phi\rightarrow\frac{\beta}{\sqrt{4\pi}}\phi$,
the conjugate momentum must rescale as
$\pi\rightarrow\frac{\sqrt{4\pi}}{\beta}\pi$.
After expressing $J^a_{\pm}$ in terms of $\phi_{\pm}$ and $\psi_{\pm}$,
and rescaling $\phi$, quantum group generators becomes
$ q^{H_1}=q^{-\frac{\beta}{\pi}(\phi(\infty)-\phi(-\infty))}$ and
\beq
 E^{\pm}_1=\sqrt{[2]/2}\sqrt{\mu/\pi}
    q^{-\frac{\beta}{2\pi}(\phi(\infty)+\phi(-\infty))}
  \int_{-\infty}^{\infty} dx
   (\psi_+e^{i\frac{4\pi}{\beta}\phi_+}
   +\psi_-e^{i\frac{4\pi}{\beta}\phi_+-i\beta\phi}),
\eeq
and similar expressions for other generators, where
$q=e^{i2\pi^2/\beta^2-i\pi/2}$. These expressions agree with the
non-local charges in the supersymmetric sine-Gordon theory\cite{KU}
up to some constant factor.
We can also show that the continuum limit
of the quantum group generators of the spin-$\frac{1}{2}$ XXZ chain
agree with the non-local charges in the sine-Gordon theory\cite{BeLe}.

We comment on other oscillator representations of the spin-1 operator:\\
(\romannumeral1)
Spin-1 version of Jordan-Wigner transformation.
Jordan-Wigner transformation for spin-$\frac{1}{2}$ case has an
advantage that there are no constraints like \eqn{const}. This is
because the Fock space on each site ($|0\rangle_n$,
$\psi^{\dagger}|0\rangle_n$, ($\psi^{\dagger 2}|0\rangle_n=0$))
agrees with the spin-$\frac{1}{2}$ representation space.
The spin-1 version is to introduce a \lq\lq parafermion" such that
its Fock space on each site is three dimensional
($|0\rangle_n$, $\psi^{\dagger}|0\rangle_n$,
$\psi^{\dagger 2}|0\rangle_n$,
($\psi^{\dagger 3}|0\rangle_n=0$)), which can be identified with
the spin-1 representation space.\\
(\romannumeral2)
Triplet real fermions.
The supersymmetric sine-Gordon Hamiltonian is expressed in terms of
the $\widehat{su}(2)_2$ currents, and the level 2 currents are
realized by a triplet of real fermions\cite{ZaFa}.
It seems natural to introduce
triplet real fermions from the beginning and write
$S^a_n=-\frac{1}{2}i\epsilon^{abc}\psi^b_n\psi^c_n$.
Real fermions on each site, however, do not allow a definite
particle picture. $S^a_n$ acts on $\psi^a_n$ as spin-1 representation
by adjoint action
$\lbrack S^a_n,\lbrack S^a_n,\psi^b_n \rbrack\rbrack=2\psi^b_n$,
and $S^a_nS^a_n=\frac{3}{4}\neq 2$, in contrast with \eqn{const}.
Nevertheless it is tempting to pursue this possibility further.
In this construction of $S^a_n$ there appear the operators $F$ and
$G^a$ which obey the OPE similar to those discussed above. The $G^a$
can be shown to satisfy the constraint \eqn{const2}. The computation
of the continuum Hamiltonian is straightforward and we get the
same form as \eqn{H} and hence the supersymmetric sine-Gordon theory.
Presumably the field theory treatment of the spin-1 chain suggested
by Tsvelik\cite{Ts} can be derived in this way.

Our derivation of the supersymmetric sine-Gordon theory as the
continuum limit of XXZ spin-1 chain is rather heuristic. However,
the fact that the connection of the quantum group generators in
the lattice theory and those in the continuum theory is correctly
obtained supports our conclusion. A rigorous proof can be
made by carrying out an analysis based on the Bethe ansatz similar
to that used to prove the equivalence of the continuum limit of
XXZ(XYZ) spin-$\frac{1}{2}$ chain and the sine-Gordon
theory\cite{LuLudN}.

We have shown that the continuum theory possesses supersymmetry.
The question arises whether the spin-1 chain has supersymmetry for
finite lattice spacing or supersymmetry emerges only in the zero
lattice spacing. This question can be answered by making a more
rigorous treatment mentioned above.

The present approach of deriving the continuum limit can be applied
to other cases of integrable spin chains:
(a) The spin-1 Hamiltonian with $b=-1$ is known to have $SU(3)$
symmetry\cite{Su}. We expect to get the affine $\widehat{su}(3)$
Toda field theory in the continuum limit. (b) For the higher spin
case we introduce $2s$ doublets of fermions to express the spin
operator. We expect to get the fractional supersymmetric sine-Gordon
theory\cite{BLMN} in the continuum.

\newpage
\section*{Acknowledgments}
\noindent
We benefited much from discussions with T. Deguchi, T. Eguchi,
I. Ichinose, P. Kulish, T. Miwa, R. Sasaki, T. Uematsu, Y. Yamada
and S.K. Yang.
We thank H. Nohara for his collaboration
in the early stage of this work.
\newpage

%
%
\end{document}